\documentclass[aps,prl,twocolumn,groupedaddress]{revtex4-1}
\usepackage{graphicx,color,times,amsmath}
\usepackage[usenames,dvipsnames]{xcolor}
\renewcommand{\vec}[1]{{\bf #1}}
\newcommand{\kB}{k_\textrm{B}}
\newcommand{\kbend}{k_\textrm{bend}}
\newcommand{\modified}[1]{{\color{red}#1}}
\begin{document}

\title{Unified analytic expressions for the entanglement length, tube diameter, and plateau modulus of polymer melts}
\author{Robert S. Hoy}
\email{rshoy@usf.edu}
\affiliation{Department of Physics, University of South Florida, Tampa, FL 33620, USA}
\author{Martin Kr\"oger}
\affiliation{Polymer Physics, ETH Z\"urich, Department of Materials,
CH--8093 Z\"urich, Switzerland}
\date{\today}

\begin{abstract}
By combining molecular dynamics simulations and topological analyses with scaling arguments, we obtain analytic expressions that quantitatively predict the entanglement length $N_e$, the plateau modulus $G$, and the tube diameter $a$ in melts that span the entire range of chain stiffnesses for which systems remain isotropic.
Our expressions resolve conflicts between previous scaling predictions for the loosely entangled [Lin-Noolandi: $G\ell_K^3/k_\textrm{B}T \sim (\ell_K/p)^3$], semiflexible [Edwards\modified{-}de Gennes:\ $G\ell_K^3/k_\textrm{B}T \sim (\ell_K/p)^2$], and tightly-entangled [Morse: $G\ell_K^3/k_\textrm{B}T \sim (\ell_K/p)^{1+\varepsilon}$] regimes, where $\ell_K$ and $p$ are respectively the Kuhn and packing lengths.
We also find that maximal entanglement (minimal $N_e$) coincides with the onset of local nematic order.
\end{abstract}

\maketitle

Individual entanglements in polymer melts and glasses are rather ethereal, delocalized objects and hence are not directly experimentally observable.
Tube models of polymer dynamics, which treat entanglements at a mean-field level, have successfully predicted many of their equilibrium and nonequilibrium properties \cite{degennes71,doi78,doi88,mcleish02,graham03}.
However, useful as these models have proven, they can tell us nothing about the behavior of individual entanglements because they represent entanglements collectively as a potential (either harmonic \cite{mcleish02,graham03} or anharmonic \cite{sussman11,sussman13,xie18}) confining a chain to its tube, and hence cannot provide a complete microscopic description of polymeric liquids.

Simulations employing  topological analysis methods based on Rubinstein and Helfand's primitive path construct \cite{rubinstein85,everaers04,kroger05,tzoumanekas06,uchida08} have provided a microscopic foundation for the tube model.
The primitive path is the shortest path a chain fixed at its ends can contract into without crossing any other chains \cite{rubinstein85,everaers04}.
The tube diameter $a$ is the characteristic extent of chains' transverse fluctuations about their primitive paths during reptation.
In a neat polymer melt,  the ``topological'' constraints limiting these transverse fluctuations are the primitive paths of the other chains \cite{rubinstein85,everaers04}.
The combination of tube theory and topological analysis has led to great advances in our understanding of polymer melt rheology \cite{li13,snijkers15}.

One major remaining gap in this understanding is that tube-theory-based formulas for the entanglement length $N_e$, tube diameter $a$, and plateau modulus $G$ of flexible, semiflexible, and stiff polymer melts are incompatible with each other.
Specifically, scaling theories for these systems assume different mechanisms of entanglement, and hence predict three different power laws for the dependence of $N_e$ on chain geometry and concentration \cite{Noolandi1987,Lin1987,edwards67,degennes74,morse01,colby90}.
Each of these power laws has been supported by both experiments \cite{fetters94,fetters99,fetters99b,hinner98} and simulations \cite{everaers04,uchida08}.
However, they are each supported only within their postulated range of validity, and quantitatively accurate theories of entanglement for systems that are intermediate between these regimes have not yet been developed.
In this Letter, we resolve this issue by presenting analytic expressions for $N_e$, $a$, and $G$ that are compatible with all three power-law-scaling regimes, and showing that they are consistent with topological analysis results for polymer melts that range from fully-flexible to nearly-stiff.

Two key quantities for characterizing the intrachain and interchain structure of polymer melts and solutions are  the Kuhn length $\ell_K$ and the packing length $p$.
$\ell_K$ can be estimated by fitting the chain statistics to the wormlike-chain model:
\begin{equation}
\displaystyle\frac{\langle R^2(n)\rangle}{n\ell_0} = \ell_K \left\{1- \displaystyle\frac{\ell_K}{2n\ell_0} \left[ 1-\exp\left(-\displaystyle\frac{2n\ell_0}{\ell_K}\right)\right]\right\},
\label{eq:lk}
\end{equation}
where $\langle R^2(n) \rangle$ is the mean-squared distance between monomers separated by chemical distance $n$ and $\ell_0$ is the backbone bond length.
$p$ is defined as $p=(\rho_c\langle R_\textrm{ee}^2\rangle)^{-1}$ \cite{fetters94}, where $\rho_c$ is the number density of chains and $\langle R_\textrm{ee}^2\rangle$ is their mean-squared end-end distance.
In neat $N$-mer melts and solutions with monomer number density $\rho$ and polymer volume fraction $\phi$, each chain occupies a volume $\Omega = N\phi/\rho = \phi/\rho_c$.
Writing this volume as $\Omega= \pi(d/2)^2 N\ell_0$ defines an effective chain diameter $d$.
The tube diameter $a$ is defined to be the end-to-end distance of a Gaussian coil of chemical length $N_e$ and effective bond length $b=\sqrt{\ell_K \ell_0}$.
Hence $N_e=(a/b)^2$ is the average number of bonds in an entangled strand, and the entanglement length is $L_e = N_e \ell_0$.

 For these systems, the plateau modulus $G$ is related to $N_e$ via $G/\kB T= \rho/N_e = \rho_e$, where $\rho_e = \rho/N_e$ is the entanglement number density \cite{doi88}.
The polymer contour length density, which was identified by Graessley and Edwards as another quantity characterizing systems' degree of entanglement \cite{graessley81}, is given by $\lambda= \rho\ell_0 =(\ell_K p)^{-1}$.
The number of Kuhn segments per chain is $N_K = Nb^2/\ell_K^2 = N/C_\infty$ (where $C_\infty=\ell_K/\ell_0$ is Flory's characteristic ratio), and the number density of Kuhn segments is $\rho_K = \rho_cN_K = \lambda/\ell_K$, implying $\ell_K/p=\rho_K \ell_K^3$.
The average number of entanglement strands per entanglement volume is $\rho_c a^3(N/N_e)=a/p$, and the average number of chains inside the volume spanned by one chain is the ``Flory number'' $n_F = \rho_c\langle R_\textrm{ee}^2\rangle^{3/2}=\sqrt{N_K} \ell_K/p$.
Thus one can write
\begin{equation}
\frac{G}{\kB T} = \frac{1}{a^2 p} = \frac{\ell_K \lambda}{a^2} = \frac{\lambda}{L_e}
\label{simpleG}
\end{equation}
and
\begin{equation}
L_e  = \frac{a^2}{\ell_K} = \frac{\lambda\kB T}{G}
\label{simpleLe}
\end{equation}
in terms of an unspecified tube diameter $a$.
Note that while for simplicity we have not included the famous factor of 4/5 (i.e.\ $G = 4\rho k_\textrm{B} T/5N_e$ \cite{larson03}) in Eqs.\ \ref{simpleG}-\ref{simpleLe}, it is trivial to do so.

Using these definitions, Everaers and collaborators have found that the dimensionless plateau modulus $G\ell_K^3/k_\textrm{B}T$ scales as $(\ell_K/p)^3$ for loosely entangled flexible-chain melts with $\ell_K \ll a$,  $(\ell_K/p)^{2}$ for $\Theta$-solutions with $\ell_K \sim a$, and $(\ell_K/p)^{7/5}$ for tightly entangled solutions of stiff chains with $\ell_K \gg a$ \cite{everaers04,uchida08}.
These scalings are broadly consistent with previous theoretical predictions and experimental results.
Fetters found that $G\ell_K^3 \sim (\ell_K/p)^3$ for a wide range of synthetic flexible polymers \cite{fetters99,fetters99b}; this scaling is predicted by the Lin-Noolandi conjecture that there are a fixed number of entangled strands per volume $a^3$ \cite{Noolandi1987,Lin1987}.
Huang \textit{et al.}\ have found that $G\sim \phi^2$ and hence $G\ell_K^3 \sim (\ell_K/p)^2$ in concentrated polystryrene solutions and melts \cite{huang15,shahid17}; this scaling is predicted by Edwards' and de Gennes' assumption that an entanglement strand corresponds to a fixed number of binary interchain contacts \cite{edwards67,degennes74}.
For isotropic solutions of stiff chains, Morse used an effective medium approximation to predict $G\ell_K^3 \sim (\ell_K/p)^{4/3}$ and a binary-collision approximation to predict  $G\ell_K^3 \sim (\ell_K/p)^{7/5}$ \cite{morse01}; experiments on tightly entangled F-actin solutions \cite{hinner98} indicate a gradual crossover from $(\ell_K/p)^{\modified{7/5}}$ to $(\ell_K/p)^{\modified{4/3}}$ scaling as $(\ell_K/p)$ increases.

In each of these scaling regimes, the dimensionless plateau modulus for well-entangled chains can be written as  \cite{graessley81}
\begin{equation}
\frac{G\ell_K^3}{\kB T} \sim \left( \frac{\ell_K}{p}\right)^\mu  = \lambda^\mu \ell_K^{2\mu},
\label{betaG}
\end{equation}
where $\mu$ is a characteristic scaling exponent.
The corresponding scaling of the tube diameter $a$ and entanglement length $L_e$ follow immediately from Eqs.\ \ref{simpleG}-\ref{betaG}:
\begin{equation}
a = \sqrt{N_e}\, b \sim \ell_K \left(\frac{\ell_K}{p}\right)^{(1-\mu)/2}
   \sim \lambda^{(1-\mu)/2}\ell_K^{2-\mu}
\label{eqa}
\end{equation}
and
\begin{equation}
L_e = \frac{\lambda \kB T}{G} \sim \ell_K \left(\frac{\ell_K}{p}\right)^{1-\mu}  \sim \lambda^{1-\mu} \ell_K^{3-2\mu},
\label{Le}
\end{equation}
where $\lambda \equiv (\ell_K p)^{-1} \sim \phi/d^2$ reflects the $\phi$-dependence.
Wang showed that $\lambda$-based expressions like Eqs.\ \ref{betaG}-\ref{Le} describe the scaling of entanglement-related quantities more accurately than earlier $C_\infty$-based expressions \cite{wang07b}.

Thus, if universal power-law dependencies of these entanglement-related quantities are assumed, the known results for flexible, semiflexible, and stiff melts (and also for melts and solutions) are incompatible with each other.
This problem was first noticed by Colby \textit{et al.} \cite{colby92} and by Fetters \textit{et al.}\ \cite{fetters94}.
Later it was analyzed in greater detail by Uchida \textit{et al.}\ \cite{uchida08}, who proposed expressions for $a/\ell_K$ and $L_e/\ell_K$ that accurately describe results for flexible and stiff systems.
More recently, it has been discussed by Milner \cite{qin14,milner20}, who suggested that (i) one has to take care about the ranges of validity of the exponent $\mu$, (ii) the Lin-Noolandi ansatz ($a \sim p$ \cite{Noolandi1987,Lin1987}) breaks down for $\ell_K \gtrsim a$, and (iii) the packing length cannot drop below the effective chain diameter, i.e.\ $p \geq d$.
The fact that Eqs.\ \ref{simpleG}-\ref{simpleLe} appear to hold for all isotropic polymer liquids suggests that universal expressions that describe all three regimes as well as the crossovers between them exist, but none have yet been developed.

To resolve this issue, we perform molecular dynamics (MD) simulations of Kremer-Grest bead-spring \cite{kremer90} polymer melts with a wide range of chain stiffnesses.
All systems are composed of $N_\textrm{ch}$ linear chains of $N$ monomers, with $N_\textrm{ch}N = 104400$.
Periodic boundary conditions are applied along all three directions of cubic simulation cells.
All monomers have mass $m$ and interact via the truncated and shifted Lennard-Jones potential $U_\textrm{LJ}(r) = 4\epsilon[(\sigma/r)^{12} - (\sigma/r)^{6} - (\sigma/r_{c})^{12} + (\sigma/r_c)^{6}]$, where $\epsilon$ is the intermonomer binding energy, $\sigma$ is the Lennard-Jones unit of length, and $r_c$ is the cutoff radius.
Covalent bonds connecting consecutive monomers along chain backbones are modeled using the FENE potential
$U_\textrm{FENE}(r) = -(k_\textrm{FENE}R_0^2/2)\ln\left[1- (r/R_0)^2\right]$, with $k_\textrm{FENE} = 30\epsilon/\sigma^2$ and $R_0=1.5\sigma$.
These choices set $\ell_0 \simeq .97\sigma$.
Angular interactions between three consecutive beads along chain backbones are modeled using the bending potential $U_b(\theta) = \kbend (1 - \cos\theta)$, where $\theta = \cos^{-1}(\hat{\vec{b}}_i\cdot \hat{\vec{b}}_{i+1})$ is the angle between consecutive bond vectors $\vec{b}_i$ and $\vec{b}_{i+1}$.
All MD simulations are performed using LAMMPS \cite{plimpton95}.

To produce melts ranging from fully-flexible to stiff, we simulate systems with $0 \leq \kbend \leq 12.5\epsilon$.
Systems are first thoroughly equilibrated at number density $\rho = 0.7\sigma^{-3}$ or $\rho = 0.85/\sigma^{-3}$ and temperature $T = 2.0\epsilon/\kB$ \footnote{We employ a higher temperature than the typical value considered in bead-spring simulations ($1.0\epsilon/\kB$ \cite{kremer90}) because the solidification temperature $T_s$ increases strongly with increasing $k_{\rm bend}$, exceeding $\epsilon/\kB$ for our stiffest $\rho = .85\sigma^{-3}$ systems.}, then run for at least one more disentanglement time ($\tau_d$) to obtain good statistics for our subsequent analyses.
We employ purely-repulsive interactions ($r_c = 2^{1/6}\sigma$) for the $\rho = 0.7\sigma^{-3}$ systems and moderate-range attractive interactions ($r_c = 2^{7/6}\sigma$) for the $\rho = 0.85/\sigma^{-3}$ systems; these parameter choices produce a sharp thermally-driven isotropic-nematic transition at $\kbend \simeq 11.5\epsilon$ for $\rho = .85\sigma^{-3}$ and a much more gradual transition for $\rho = .7\sigma^{-3}$.
Equilibrating the stiff-chain isotropic and nematic systems requires a nonstandard protocol \cite{SI}.
The attractive part of the interactions is not necessary to reproduce the results discussed below; the key difference between the $\rho = .7\sigma^{-3}$ and $\rho = .85\sigma^{-3}$ systems is their different $\lambda$.

We perform topological analyses of statistically independent equilibrated-melt snapshots using the Z1 algorithm \cite{kroger05}.
The code returns each chain's number of kinks (entanglements) $Z$.
We calculate $N_e$ using the ideal ``M-kink'' estimator derived in Ref.\ \cite{hoy09}: $N_e^{-1}=d\langle Z\rangle/dN$,
where the average is taken over all chains and statistically independent melt configurations having the same $\kbend$.
This estimator can be employed here because $\langle Z\rangle$ is approximately linear in $N$ for all $\kbend$ over the chain length range ($50 \leq N \leq 100$) we study.

\begin{figure}[h]
\includegraphics[width=2.6in]{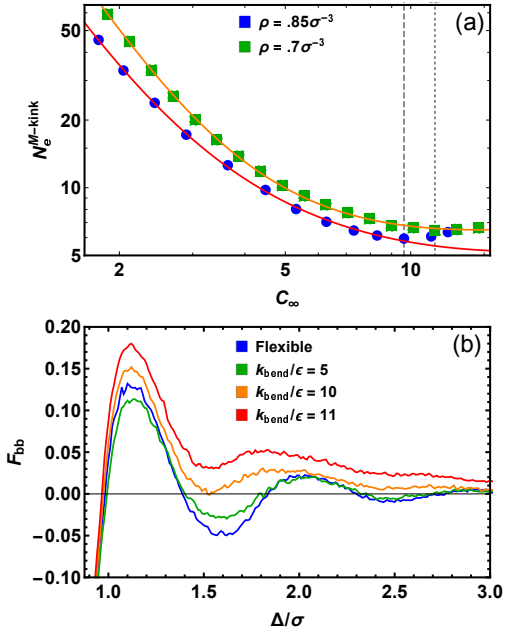}
\caption{Relation of entanglement to local structure.  Panel (a):\ $N_e$ vs. $C_\infty=\ell_K/\ell_0$ for our Kremer-Grest melts \cite{SI}. Z1 data (blue and green symbols) and the analytic expression given in Eq.\ \ref{eq:NeofCinf} (red and orange curves).  The dashed and dotted vertical lines indicate $C_\infty^*(\rho)$ for $\rho = .85\sigma^{-3}$ and $\rho = .7\sigma^{-3}$ systems.  Panel (b): $F_\textrm{bb}(\Delta)$ for selected $\kbend$ for the $\rho = .85\sigma^{-3}$ systems.}
\label{fig:1}
\end{figure}

Figure \ref{fig:1}(a) shows Z1 results for all systems.
For both $\rho$, $N_e$ decreases sharply with $C_\infty$ at small $C_\infty$, decreases progressively less sharply as $C_\infty$ increases, and passes through a minimum at $C_\infty = C_\infty^*(\rho)$.
We find that the $C_\infty$-dependence of the measured $N_e$ for $C_\infty \leq C_\infty^*(\rho)$ is quantitatively captured (for a given $\lambda$) by
a sum of three terms:
\begin{equation}
N_e(C_\infty) =  \alpha C_\infty^{-3} + \beta C_\infty^{-1} + \gamma C_\infty^{1-2\varepsilon}.
\label{eq:NeofCinf}
\end{equation}
Our results are consistent with $.25 \lesssim \varepsilon \lesssim .4$; below, we will assume $\varepsilon = 1/3$.
Values for $C_\infty^*(\rho)$ and all parameters in Eq.\ \ref{eq:NeofCinf} are given in Table \ref{tab:params}.

\begin{table}[h]
\caption{Stiffness at maximal entanglement and parameter values for Eq.\ \ref{eq:NeofCinf} for dense Kremer-Grest melts at $k_\textrm{B}T = 2.0\epsilon$.}
\begin{ruledtabular}
\begin{tabular}{cccccc}
$\rho\sigma^3$ & $\lambda\sigma^2$ & $\alpha$ & $\beta$ & $\gamma$ & $C_\infty^*$ \\
.85 & .817  & 171$\pm$3 & 23.8$\pm$0.9 & 1.47$\pm$.09  & 9.63\\
.7 &  .673  & 303$\pm$3 & 20.1$\pm$0.9 & 2.05$\pm$.07 & 11.43
\end{tabular}
\end{ruledtabular}
\label{tab:params}
\end{table}

In light of the above scaling arguments (which imply $N_e \sim (\rho\ell_0^3)^{1-\mu} C_\infty^{3-2\mu}$ \cite{graessley81}), the contributions to the measured $N_e$ with numerical prefactors $\alpha$, $\beta$, and $\gamma$ respectively correspond to $\mu=3$, $\mu=2$, and $\mu= 1 + \varepsilon = 4/3$.
Since the entanglement length $N_e^\textrm{PPA}$ measured by PPA analyses is roughly $3N_e^Z/2$ \cite{hoy09,everaers12}, the minima in Fig.\ref{fig:1}(a) occur at $C_\infty \simeq N_e^\textrm{PPA}$.
This coincides with Milner's operational definition \cite{milner20} of the crossover between the semiflexible and stiff-chain regimes:\ $N_e \simeq C_\infty$ and $a \simeq \ell_K$.
The physical interpretation of this crossover is that it occurs when entangled strands and Kuhn segments coincide.

One potential model-agnostic, microstructural reason for the nonmonotonicity of $N_e(C_\infty)$ is that local nematic order reduces entanglement and hence the minima in $N_e$ correspond to the onset of such order.
Figure \ref{fig:1}(b) shows results for the radially symmetric bond-orientational correlation function
\begin{equation}
F_\textrm{bb}(\Delta) = \left\langle \left|\hat{\vec{b}}_i(\vec{R}_i)\cdot\hat{\vec{b}}_j(\vec{R}_i + \vec{\Delta}_{ij})\right|\right\rangle - \displaystyle\frac{1}{2},
\label{eq:fbb}
\end{equation}
for selected $\rho = .85\sigma^{-3}$ systems
\footnote{Here $\vec{R}_i$ indicates the midpoint of the bond $\vec{b}_i,$ $\vec{\Delta}_{ij} = \vec{R}_j - \vec{R}_i$, and the brackets denote averages over all $j > i$ for which $|\vec{\Delta}_{ij}|=\Delta$.  Only bonds belonging to different chains are included in the averages.}.
$F_\textrm{bb}(\Delta)$ is a sensitive measure of scale-dependent nematic order that is positive when bond vectors separated by a midpoint--midpoint distance $\Delta$ are correlated, and zero when they are uncorrelated.
The data show that the value of the first minimum in $F_\textrm{bb}(r)$, $F_\textrm{bb}^\textrm{min}$, is negative for $\kbend/\epsilon < 10$ and positive for $\kbend/\epsilon > 10$, i.e.\ $F_\textrm{bb}^\textrm{min}$ is negative for $C_\infty < C_\infty^*$ and positive for $C_\infty > C_\infty^*$.
Positive $F_\textrm{bb}^\textrm{min}$ indicate that chains are locally aligned; we find that when $F_\textrm{bb}^\textrm{min} > 0$, chains also remain aligned out to considerably larger $\Delta$.
Analogous results hold for the $\rho = .7\sigma^{-3}$ systems.
Strictly speaking, Eqs.\ \ref{simpleG}-\ref{Le} hold only for isotropic melts and solutions, which lack even local nematic order.
For this reason, we included only data for $C_\infty < C_\infty^*(\rho)$ in the fits of the Z1 data to Eq.\ \ref{eq:NeofCinf}.

Uchida \textit{et al.}\ showed that primitive path network-viscoelastic property relations are system-independent, i.e.\ the relation $G = \rho k_\textrm{B} T/N_e$ remains valid over the \textit{entire} range of chain stiffnesses for which systems remain isotropic \cite{uchida08}.
The fact that $N_e$ can be expressed as a sum of contributions from flexible-, semiflexible-, and stiff-chain entanglement mechanisms (i.e.\ Eq.\ \ref{eq:NeofCinf}) that is quantitatively accurate for all $C_\infty < C_\infty^*(\rho)$ suggests that corresponding expressions can be found for the tube diameter and plateau modulus.
We now attempt to do so.

\begin{figure*}
\includegraphics[width=7in]{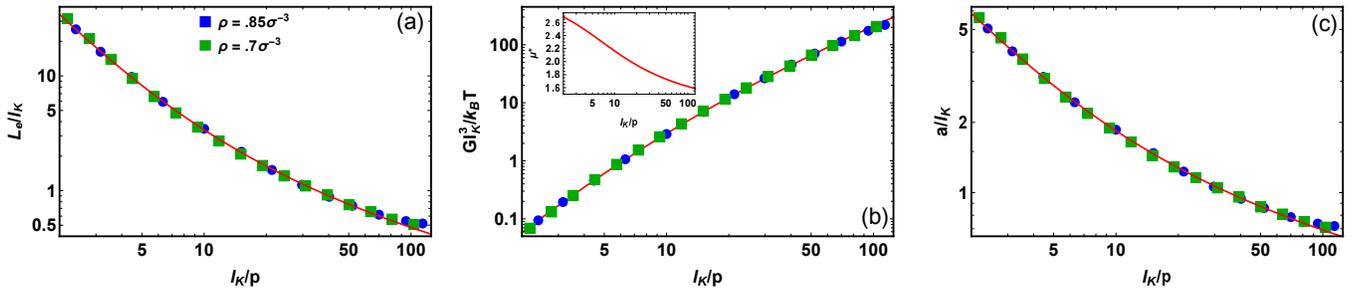}
\caption{$L_e/\ell_K$, $G\ell_K^3/k_\textrm{B}T$ and $a/\ell_K$ vs. $\ell_K/p$ for our Kremer-Grest melts.  Blue and green symbols in panels (a-c) respectively show results for $L_e/\ell_K \equiv N_e^\textrm{M-kink}/C_\infty$, $G \ell_K^3/k_\textrm{B}T \equiv \rho\ell_0^3 C_\infty^3/N_e^\textrm{M-kink}$ and $a/\ell_K \equiv \sqrt{N_e^\textrm{M-kink}/C_\infty}$ vs.\ $\ell_K/p \equiv \rho\ell_0^3 C_\infty^2$, where $C_\infty$ is calculated from systems' chain statistics (Eq.\ \ref{eq:lk}).
Orange curves show Eqs.\ \ref{LeApproximant}-\ref{eq:aanalyt} with the $\{ c_i \}$ given above and $\varepsilon = 1/3$. The inset to panel (b) shows the local scaling exponent $\mu^*$ calculated from Eq.\ \ref{eq:derivedG} using Eq.\ \ref{eq:mustar}.}
\label{fig:2}
\end{figure*}

The identity $\ell_K/p = \rho \ell_0^3 C_\infty^2$ leads to another analytic expression for the reduced entanglement length that is equivalent to Eq.\ \ref{eq:NeofCinf}:
\begin{equation}
\frac{N_e}{C_\infty} \equiv \frac{L_e}{\ell_K} = c_1\left(\frac{\ell_K}{p}\right)^{-2} + c_2\left(\frac{\ell_K}{p}\right)^{-1}  + c_3 \left(\frac{\ell_K}{p}\right)^{-\varepsilon}
\label{LeApproximant}
\end{equation}
Plugging the above values of ($\alpha, \beta, \gamma$) into Eq.\ \ref{LeApproximant} yields the values of its $c$-coefficients:\  for $\rho = .85\sigma^{-3}$, $c_1 = \alpha (\rho \ell_0)^2 \simeq 97.5$, $c_2 = \beta(\rho \ell_0^3) \simeq 17.9$, and $c_3 = \gamma (\rho \ell_0^3)^\varepsilon \simeq 1.34$.
Thus the dimensionless plateau modulus is
\begin{equation}
\displaystyle\frac{G \ell_K^3}{k_\textrm{B} T} = \left[c_1\left(\frac{\ell_K}{p}\right)^{-3} + c_2\left(\frac{\ell_K}{p}\right)^{-2} + c_3\left(\frac{\ell_K}{p}\right)^{-(1+\varepsilon)}\right]^{-1}.
\label{eq:derivedG}
\end{equation}
The identity  $a = \ell_K (L_e/\ell_K)^{1/2}$ leads to an analytic expression for the reduced tube diameter:
\begin{equation}
\frac{a}{\ell_K} = \sqrt{ c_1\left(\frac{\ell_K}{p}\right)^{-2} +c_2\left(\frac{\ell_K}{p}\right)^{-1}  +c_3 \left(\frac{\ell_K}{p}\right)^{-\varepsilon} }.
\label{eq:aanalyt}
\end{equation}
The validity of Eqs.\ \ref{LeApproximant}-\ref{eq:aanalyt} should be independent of chain stiffness and thickness (i.e.\ ``chemistry'') as well as chain length.
In particular, the $\{ c_i \}$ in Eqs.\ \ref{LeApproximant}-\ref{eq:aanalyt} should be chemistry-independent because (as shown by Fetters \cite{fetters94,fetters99,fetters99b}) the chemistry-dependence is contained in $\ell_K$ and $p$.

Figure \ref{fig:2} compares the predictions of Eqs.\ \ref{LeApproximant}-\ref{eq:aanalyt} to MD/Z results for these quantities.
Remarkably, results for all systems with $C_\infty \leq C_\infty^*(\rho)$ collapse onto single master curves even though the $\{ c_i \}$ values in Eqs.\ \ref{LeApproximant}-\ref{eq:aanalyt} were calculated using only the $\rho = .85\sigma^{-3}$ data.
These collapses are nontrivial because only the values of $\rho$, $\ell_0$, $C_\infty$, and $N_e^\textrm{M-kink}$ taken from the MD and Z simulations were used to evaluate the estimates of $L_e/\ell_K$, $G\ell_K^3/k_\textrm{B} T$, and $a/\ell_K$; no additional fitting was performed.
It is evident that Eqs.\ \ref{eq:NeofCinf} and \ref{LeApproximant} quantitatively capture the crossovers between the $\mu = 3$, $\mu = 2$, and $\mu = (1+\varepsilon)$ scaling regimes.
Consequently, Eqs.\ \ref{eq:derivedG}-\ref{eq:aanalyt} also quantitatively capture these crossovers.
The intermediate-stiffness ($\mu = 2$) regime can be roughly defined by the criteria $\alpha/\beta \le C_\infty^2\le (\beta/\gamma)^{1/(1-\varepsilon)}$ [i.e.\  $2.7 \leq C_\infty \leq 8.1$] or equivalently  $c_1/c_2 \le \ell_K/p \le (c_2/c_3)^{1/(1-\varepsilon)}$ [i.e.\ $5.4 \leq \ell_K/p \leq 48$], or alternatively by plotting the local scaling exponent \cite{SI}
\begin{equation}
\mu^*\left(\frac{\ell_K}{p} \right) = \displaystyle\frac{\partial \left[ \ln(G\ell_K^3/k_\textrm{B} T) \right]}{\partial \left[ \ln(\ell_K/p) \right]} = 3 + \ell_K \frac{\partial \ln G}{\partial \ell_K}.
\label{eq:mustar}
\end{equation}
As shown in the inset to panel (b), $\mu^*$ decreases rather smoothly from $\sim 8/3$ to $\sim 8/5$ over the range of ($\ell_K/p$) considered here.
We remain agnostic about the precise value of $\varepsilon$, but we have found no evidence that it is outside the range $1/3 \leq \varepsilon \leq 2/5$ specified by Morse \cite{morse01}.

Equations \ref{LeApproximant}-\ref{eq:aanalyt} are nonstandard and require interpretation.
Recall that the entanglement length $N_e$ can be regarded as a dimensionless elastic compliance:\ $N_e = \rho k_\textrm{B} T/G$.
One physically plausible interpretation of Eqs.\ \ref{eq:NeofCinf} and \ref{LeApproximant} is that $N_e$ is, in general, a sum of three elastic compliances that add in series, where the compliances $\alpha C_\infty^{-3}$, $\beta C_\infty^{-1}$ and $\gamma C_\infty^{1-2\varepsilon}$ respectively represent the contributions of $\mu = 3$-, $2$-, and ($1+\varepsilon$)-entanglement mechanisms to $G$.
If this were true, the contributions of their associated elastic moduli $G_3\ell_K^3/k_\textrm{B}T = c_1^{-1}(l_K/p)^3$, $G_2\ell_K^3/k_\textrm{B}T = c_2^{-1}(l_K/p)^2$ and $G_{1+\varepsilon} \ell_K^3/k_\textrm{B}T = c_3^{-1}(l_K/p)^{1+\varepsilon}$ to the overall modulus would be given by the equation
\begin{equation}
G = \left[ \frac{1}{G_3} + \frac{1}{G_2} + \frac{1}{G_{1+\varepsilon}} \right]^{-1}.
\label{eq:parallelG}
\end{equation}
Eq.\ \ref{eq:derivedG} indeed has this form.
Our interpretation of Eqs.\ \ref{LeApproximant}-\ref{eq:aanalyt} is that the previously identified \cite{Noolandi1987,Lin1987,edwards67,degennes74,morse01} contributions of flexible-, semiflexible-, and stiff-chain-entanglement mechanisms to $G$ can be \textit{mathematically} represented as elastic springs arranged in series.
In this picture, $G$ is dominated by the smallest of $\{ G_{1+\varepsilon}, G_2, G_3 \}$, and the previously identified scaling regimes are recovered when one of the $G_i$ is small compared to the other two.
However, we believe that Eqs.\ \ref{LeApproximant}-\ref{eq:aanalyt} do not in fact represent three mechanisms of network elasticity that act independently; after all, there is no test that can determine whether a given entanglement is flexible, semiflexible, or stiff.
Instead, these equations are simply functional forms that capture how the connections between a polymer melt's local structure and its degree of entanglement vary with chain stiffness and contour length density.

In conclusion, we have in this Letter derived unified analytic expressions for the reduced entanglement length, plateau modulus, and tube diameter of polymer melts that appear to be valid over the entire range of chain stiffnesses for which systems remain isotropic.
Our results are compatible with previous scaling theories \cite{Noolandi1987,Lin1987,edwards67,degennes74,morse01,uchida08,milner20} as well as experimental results \cite{fetters94,fetters99,fetters99b,Ruymbeke2019,hinner98} for systems that span this range, but go beyond previous results by capturing the crossovers between the $\mu = 3$-, $2$-, and ($1+\varepsilon$)-scaling regimes and providing the relevant numerical prefactors.
In particular, while Uchida \textit{et. al.} derived crossover expressions \cite{uchida08} for $L_e/\ell_K$, $G\ell_K^3/k_\textrm{B} T$ and $a/\ell_K$ that are comparable to Eqs.\ \ref{LeApproximant}-\ref{eq:aanalyt} and fit results for flexible melts and stiff solutions very well, \modified{t}heir expressions lacked the $\mu = 2$ terms (i.e.\ the $c_2$-terms) and hence did not accurately capture results for semiflexible melts \cite{SI}.

We emphasize that Eqs.\  \ref{LeApproximant}-\ref{eq:aanalyt} are intended \textit{only} for dense melts and very concentrated solutions.
They must eventually break down as $\phi$ decreases because they cannot capture the crossover to the Colby-Rubinstein  ($\mu = 7/3$) scaling \cite{colby90} observed in semidilute solutions \cite{colby91,huang15,inoue02,shahid17}.
Future work will consider solutions and will examine how expressions like Eqs.\  \ref{LeApproximant}-\ref{eq:aanalyt} break down as $\phi$ decreases.

Kenneth S. Schweizer and Scott T. Milner provided helpful discussions.
This material is based upon work supported by the National Science Foundation under Grant DMR-1555242 and by the Swiss National Science Foundation through Grant 200021L\_185052.


%

\end{document}